\documentclass[10pt, aps,prl,twocolumn,superscriptaddress,preprintnumbers,
amsmath, 
floatfix,
longbibliography,
showpacs 
]{revtex4-1}
\usepackage[T1]{fontenc}
\usepackage[utf8x]{inputenc} 
\usepackage{adjustbox}          
\usepackage[caption=false]{subfig}
\usepackage{url}
\usepackage{color}
\usepackage{float}
\usepackage{hyperref}
\usepackage{cleveref}
\usepackage{longtable}
\usepackage{amsfonts}
\usepackage{wrapfig}

\newcommand{\beq}{\begin{equation}}
\newcommand{\eeq}{\end{equation}}
\newcommand{\bea}{\begin{eqnarray}}
\newcommand{\eea}{\end{eqnarray}}

\hypersetup{pdftex,colorlinks=true, linkcolor=blue, citecolor=blue,urlcolor=blue, bookmarksnumbered=true, bookmarksopen=true}

\begin{document}

\title{Dynamics of Fragmented Condensates and Macroscopic Entanglement}
  
\author{Aurel Bulgac}\email{bulgac@uw.edu}%
\author{Shi Jin} 

\affiliation{
Department of Physics,%
  University of Washington, Seattle, Washington 98195--1560, USA}

\date{\today}

\begin{abstract}

  The relative phase of the order parameters in the collision of two
condensates can influence the outcome of their collision in the case
of weak coupling. With increasing interaction strength however, the
initially independent phases of the two order parameters in the
colliding partners quickly become phase locked, as the strong coupling
favors an overall phase rigidity of the entire condensate, and upon
their separation the emerging superfluid fragments become entangled.

\end{abstract}

\preprint{NT@UW-17-02}
\pacs{}

\maketitle

Since the gauge symmetry is spontaneously broken in superfluids, it is
reasonable to wonder under what conditions the relative phase of two
superfluids is physically relevant. The Josephson
effect~\cite{Josephson_rmp1964, Josephson_rmp1974}, experiments with
cold Bose or Fermi atoms~\cite{shin:2004, Yefsah:2013,
Zwierlein_prl2014, Zwierlein_prl2016, Bulgac_prl2014,
Wlazlowski_pra2015}, and the superfluid fragments emerging from
nuclear fission~\cite{Bulgac:2017, Bulgac:2017c, Bulgac:2017d}, are
just a few examples where that is the case.  As we will discuss here,
there are other situations when one would however expect that the
relative phase of two condensates is physically irrelevant. However,
the emerging overall picture of the role of the relative phase of two
condensates appears to be more complex than envisaged so far.
Recently Magierski, Sekizawa, and Wlaz{\l}owski
(MSW)~\cite{Magierski:2016} reported on a rather surprising
observation concerning the role the pairing field plays in the
collisions of two heavy-ions at energies near the Coulomb barrier. MSW
observed a very strong dependence of the properties of the emerging
fragments on the relative phase of the pairing condensates in the
initial colliding nuclei. In a somewhat related study of
$^{20}$O+$^{20}$O~\cite{Hashimoto:2016}, the reported effect was
rather weak, a result confirmed in the similar case of
$^{44}$Ca+$^{44}$Ca~\cite{Sekizawa:2017}, due to the small number of
nucleons above the closed shell.  The amplitude of the pairing field
$\Delta$ in nuclei is of the order of 1 MeV, which is significantly
smaller than the magnitude of the normal single particle field, which
is of the order of 50 MeV. The character of the nuclear pairing
correlations is recognized in literature of being of the
Bardeen-Cooper-Schriefer (BCS) type~\cite{Bardeen:1957}, a theory
which describes weak coupling pairing with Cooper pairs with sizes
significantly larger than the average separation between fermions.
The gain in binding energy due to pairing correlations, called
condensation energy $E_{cond}=-N(0)|\Delta|^2/2$, can hardly be
greater than perhaps a few MeVs.  MSW report however that in the
collision of $^{240}$Pu on $^{240}$Pu near the Coulomb barrier pairing
effects can lead to changes in the total kinetic energy of the
emerging fragments of up to 20 MeV and that the apparent height of the
fusion barrier could be changed by 10 MeV or even more.  These
dramatic changes, with an energy significantly higher than the
magnitude of the total pairing condensation energy, were correlated by
MSW with the relative phase of the pairing fields in the two colliding
partners prior to collisions.

The gauge symmetry breaking bears similarity with the rotational
symmetry breaking in case of deformed nuclei, when their relative
orientations plays a noticeable role in heavy-ion fusion reactions and
various decays.  The MSW results, obtained by solving the
time-dependent density functional theory (TDDFT) equations, can be
reproduced semi-quantitatively using a simple Ginzburg-Landau (GL)
approach~\cite{Ginzburg:1950, Ginzburg_rmp2004}, or the formally
equivalent static Gross-Pitaevskii (GP) description~\cite{Gross:1961,
Pitaevskii:1961}.  When the two nuclei touch the phase of the
condensate can change across the contact region, as in a domain wall,
in a manner superficially similar to the tunneling current in a
Josephson junction~\cite{Josephson_rmp1964, Josephson_rmp1974}, albeit
in the absence of a barrier.
 
In the presence of pairing correlations the ground state of a nucleus
is a Bose-Einstein condensate (BEC) of Cooper pairs, which in theory
is accurately described in the grand canonical ensemble, where only
the average particle number is specified. The phase of the order
parameter $\hat{\phi}$ is conjugate to the particle number $\hat{N}$,
and thus in a system with well defined particle number the phase is
undefined~\cite{Anderson_rmp1966, Carruthers_rmp1968}.  However, as
Anderson points out~\cite{Anderson:1986}: in a bucket of liquid helium
below the $\lambda$-point ``$\phi$ has become a classical variable,
...  any future experiment will be interpretable as though $\phi$ was
fixed.''  This is also the prevalent approach in describing nuclei
with well defined pairing correlations, when the effect of particle
projection is small.  One can thus reasonably ask, a common question
in condensed matter physics: ``Can a nucleus have a well defined phase
of the condensate with respect to another nucleus?''  Since the total
wave function of the two nuclei prior to their interaction is merely a
product of two independent wave functions, one would expect that the
interaction between two nuclei cannot depend on the phases of each
initial wave functions.  A (relative separation) coordinate dependence
of the phase of the pairing field indicates the presence of a current.
The phases of the pairing fields can be changed by arbitrary and
independent gauge transformations in each partner prior to the moment
the two nuclei touch and thus one can generate a phase gradient in the
``neck.''  An objection raised by G.F. Bertsch in discussions was that
initial nuclei have well defined proton and neutron numbers, unlike
the anomalous densities which are the central objects in a DFT
approach, and the phase of the wave function of each nucleus prior to
the collision should be physically irrelevant.  Clearly a similar
argument would not be accepted in case of deformed nuclei, as a number
of observables are impacted ($\alpha$-decay penetrability, heavy-ion
fusion cross sections, etc.)  This kind of argumentation began at
the inception of quantum mechanics and many have wondered about
similar problems, see Anderson's talk~\cite{Anderson:1986} and the
follow-up spirited discussion. As Anderson writes: ``if the
experimenter now cools down two entirely different, non-communicating
buckets of liquid helium from $T>T_\lambda\rightarrow T\approx 0$,
... upon opening an orifice between the two, would see initially with
equal probability any {\it fixed} value of the phase difference, and
thereafter no experiment he tried could recover the components of the
wave-function which started out with different relative phases. He
would not see {\it zero} interference current, ...''  This situation
corresponds theoretically to a fragmented
condensate~\cite{Leggett:2006} and the inability of the experimenter
to recover the initial state is due the fact that the two buckets
became macroscopically entangled after being in contact for some time.
Macroscopic entanglement of up to hundreds to millions of particles
have been put in evidence experimentally~\cite{Iskhakov:2012,
Behbood:2014, McConnell:2015, Froewis:2017, Zarkeshian:2017}. It is
crucial to recognize that there are two qualitative steps in
Anderson's gedanken experiment, the creation of the initial state and
the subsequent emergence of the final state. This is also the
situation in the MSW simulations and the natural question arises, why
these authors did not observe the outcome conjectured in Anderson's
gedanken experiment, as the outcome of their collisions showed a
strong dependence on the initial relative phase of the condensates,
unlike what Anderson conjectured. We are not aware however of any
experiments in which the dependence on the strength of the coupling on
the outcome of a collision and of the entanglement have been studied.

There is however another qualitatively different situation, relevant
to experiments performed in cold gases~\cite{shin:2004, Yefsah:2013,
  Zwierlein_prl2014, Zwierlein_prl2016, Bulgac_prl2014,
 Wlazlowski_pra2015} or to superfluid fragments emerging from
nuclear fission~\cite{Bulgac:2017, Bulgac:2017c, Bulgac:2017d}.  This
happens when one cools down a bucket of helium from above the
$\lambda$-transition, and subsequently separates it into two parts
kept always close to $T\approx 0$ and reunites them after they had
different histories, and the two parts remain macroscopically
entangled at all times~\cite{Anderson:1986}.  In this situation the
relative phase of the two buckets is always rather well defined, but
the particle numbers in the two buckets are not.  (We will not discuss
here the role the phase diffusion can play.)

There will definitely be increasingly more studies of colliding
superfluid nuclei and other systems in the future performed within the
only practical microscopic framework available so far, the TDDFT. A
correct interpretation of such numerical simulation results and a
correct method to evaluate observables are stringent elements of our
theoretical tools, tools which are still not yet ascertained.  Nuclei
contain many particles, are essentially macroscopic objects, and as
Anderson has also noted~\cite{Anderson:1986}: ``... the central
problem of measurement theory is not the quantum mechanics of atoms,
which is simple and easy, but the fact that macroscopic everyday
objects are very difficult indeed for the quantum theory to deal with
properly.''  Many properties of nuclei (liquid drop mass formula,
surface tension, compressibility, symmetry energy, hydrodynamics,
collective motion, rotation, symmetry breaking, transport
coefficients, etc.) can be and are often treated quite accurately
using concepts characteristic for macroscopic systems.

In order to shed light on MSW's very startling observation, that the
relative phase of the pairing fields in two colliding nuclei can have
a dramatic role in the collision process, we will turn at first to a
simpler system, in which the role of the relative phase of two
condensate can be easily studied.  In the presence of pairing
correlations nuclei can be treated as a BEC of interacting Cooper
pairs, as in the case of electrons in
superconductors~\cite{Bardeen:1957}, and the total wave function can
be represented as an anti-symmetrized product of Cooper pair wave
functions.  In the case of a weakly interacting Bose system at zero
temperature a GP equation is extremely
accurate~\cite{Dalfovo_rmp1999}.  In the GP approximation a boson
field operator $\hat{\psi}({\bf r})$ is replaced with its
non-vanishing average $ \psi( {\bf r} )= \langle 0 | \hat{\psi} ( {\bf
r} ) |0\rangle$ [a classic example of $U(1)$ broken gauge symmetry]
and the accuracy of the approximation is of order $\sim 1/\sqrt{N}$,
where $N$ is the total number of bosons.  A BCS fermionic condensate
is a system of weakly interacting Cooper pairs/bosons and
qualitatively a GP equation is appropriate and has been used in the
past numerous times.  The weakness of the interaction is typically
characterized by the ratio of the pairing gap to the Fermi energy
$\Delta/\varepsilon_F\ll 1$. In the weak coupling limit, all Cooper
pairs have a zero momentum, as in a BEC.

Typical BEC systems have all particles in one cloud and the one-body
density matrix acquires the form
\beq \rho({\bf r}_1, {\bf r}_2) =
\langle 0| \hat{\psi}^\dagger ({\bf r}_1)\hat{\psi} ({\bf
  r}_2)|0\rangle \Rightarrow n_0 \psi^*( {\bf r}_1 )\psi ( {\bf r}_2
), \nonumber 
\eeq 
when $ | {\bf r}_1 - {\bf r}_2 | \rightarrow \infty$, and there is
only one eigenvector with a macroscopic eigenvalue $n_0={\cal O}(N)$,
a situation known as the off-diagonal long-range order
(ODLRO)~\cite{Ginzburg:1950,Penrose:1951,Penrose:1956,Yang_rmp1962}.
It is possible to have a fragmented BEC system~\cite{Leggett:2006},
when two or more eigenvalues of the one-body density matrix $\rho({\bf
r}_1, {\bf r}_2)$ are macroscopically large.  This is the case of two
BEC clouds with particle numbers $N_1$ and $N_2$ in two spatially well
separated potential traps $V_k({\bf r})$, $\int d^3 {\bf r} \left |
\psi_k ( {\bf r},t) \right |^2=N_k$, $k=1,2$,
\bea &&i\hbar\dot{\psi}_k ( {\bf r}, t) = -\frac{\hbar^2}{2m}
\triangle \psi_k({\bf r},t)+ V_k({\bf r})\psi_k({\bf r},t) \label{eq:1B} \\
&&+g\left | \psi_k({\bf r},t)\right |^2\psi_k({\bf r},t) =
\mu_k\psi_k({\bf r}, t)=\mu_k\phi_k({\bf r})
e^{-i\mu_kt/\hbar}.  \nonumber 
\eea 
Let us consider now this fragmented BEC condensate, when their
initially spatially well separated trapping potentials are moving
towards each other, and their combined wave function at times before
the two clouds come into contact is naturally given by
\bea
&& \Psi({\bf r},t) =  \psi_1({\bf r},t)+ e^{i\alpha} \psi_2({\bf r},t), \label{eq:NB}\\
&& \psi_k({\bf r},t) =\phi_{k}({\bf r}-{\bf r}_k - {\bf u}_kt)
e^{i (m{\bf u}_k\cdot{\bf r} - \mu_k t/\hbar - m{\bf u}^2_kt/2)/\hbar } \label{eq:1b}, 
  \eea 
with $V_k({\bf r}) \rightarrow U_k({\bf r},t) = V_k ({\bf r}-{\bf
r}_k-{\bf u}_kt)$ and $ e^{i\alpha} $ arbitrary.  Using $\Psi({\bf
r},t)$ one can construct a coherent state $\exp[\tau \int d^3{\bf
r}\Psi({\bf r},t)\hat{\psi}^\dagger({\bf r}) ]\, |0\rangle$ and the
fragmented BEC state is obtained only after a specific particle
projection is performed, see the discussion below and in connection
with Eq.~(\ref{eq:2N}) and the Supplemental Online Material
(SOM)~\cite{Supplement}.  We will assume that the velocities ${\bf
u}_k$ are significantly smaller in magnitude than the speed of
sound~\cite{Dalfovo_rmp1999} $c=\sqrt{g|\Psi({\bf r},t)|^2/m}$
evaluated in the central part of the cloud, and therefore
superfluidity is not endangered.  At all times this combined wave
function satisfies the time-dependent GP equation with $U({\bf
r},t)=U_1({\bf r},t)+U_2({\bf r},t)$,
\beq \label{eq:GP}
i\hbar\dot{\Psi}( {\bf r}, t) = \left [ -\frac{\hbar^2}{2m}\triangle
  +g\left | \Psi({\bf r},t)\right |^2+ U({\bf r},t)\right ]\Psi({\bf
  r},t). \nonumber 
\eeq 
Before contact each component of the total wave function
$\psi_{k}({\bf r},t)$, see Eq. (\ref{eq:1B}), satisfies its own
time-dependent GP equation (\ref{eq:1B}) with $V_k({\bf r})
\rightarrow U_k({\bf r},t)$.  The arbitrary phase $\exp(i\alpha)$ can
arguably influence the dynamics if $g\neq0$.  This is the phase in one
of the two cases of liquid helium buckets discussed by
Anderson~\cite{Anderson:1986}.  Unlike the overall phase of the
many-body wave function, this phase cannot be removed now, similarly
to the relative orientation of two colliding deformed nuclei.  In the
case of two separated condensates the overall order parameter is the
sum of the two separated order parameters, similarly to magnetization
for example.  (The action of the magnetic field on the spin coordinate
of a fermion is formally identical to the action of the pairing field
on the two components of the fermionic
quasiparticle~\cite{Anderson:1958}.)  Magnetization is created by
electric currents and magnetic moments, and when one brings two
magnets into proximity, the two magnetic fields add up, even though
the many-body electron wave functions for the two separated magnets
are multiplied to each other. As in the case of a magnetic field,
where the relative orientation of two magnetic fields is important,
and in the case of the complex pairing field the relative phase of the
two fields is important, as is in the case of Josephson junctions too.
This relative phase is also arbitrary, but this relative phase can be
controlled in some instances. In the vicinity of an isolated cloud one
can apply for a finite interval of time a constant potential over the
isolated cloud, a procedure performed in the case of cold atoms in
experiments, equivalent to performing a local gauge transformation,
and thus one can change the relative phase of two
clouds~\cite{Yefsah:2013, Zwierlein_prl2014, Zwierlein_prl2016,
Bulgac_prl2014, Wlazlowski_pra2015}.
 
By analyzing both the GP equation, see SOM~\cite{Supplement}, and the
collision of superfluid nuclei we arrived at a totally unexpected and
surprising result, that the strength of the interaction $g$ plays a
qualitative role in the dynamics.  By increasing the strength of the
interaction from zero (corresponding to the case of non-interacting
bosons or absence of pairing correlations in nuclei) to a relatively
large value, the character of the collision changes dramatically, but
in a continuous manner.

\begin{figure}
\includegraphics[width=0.75\columnwidth]{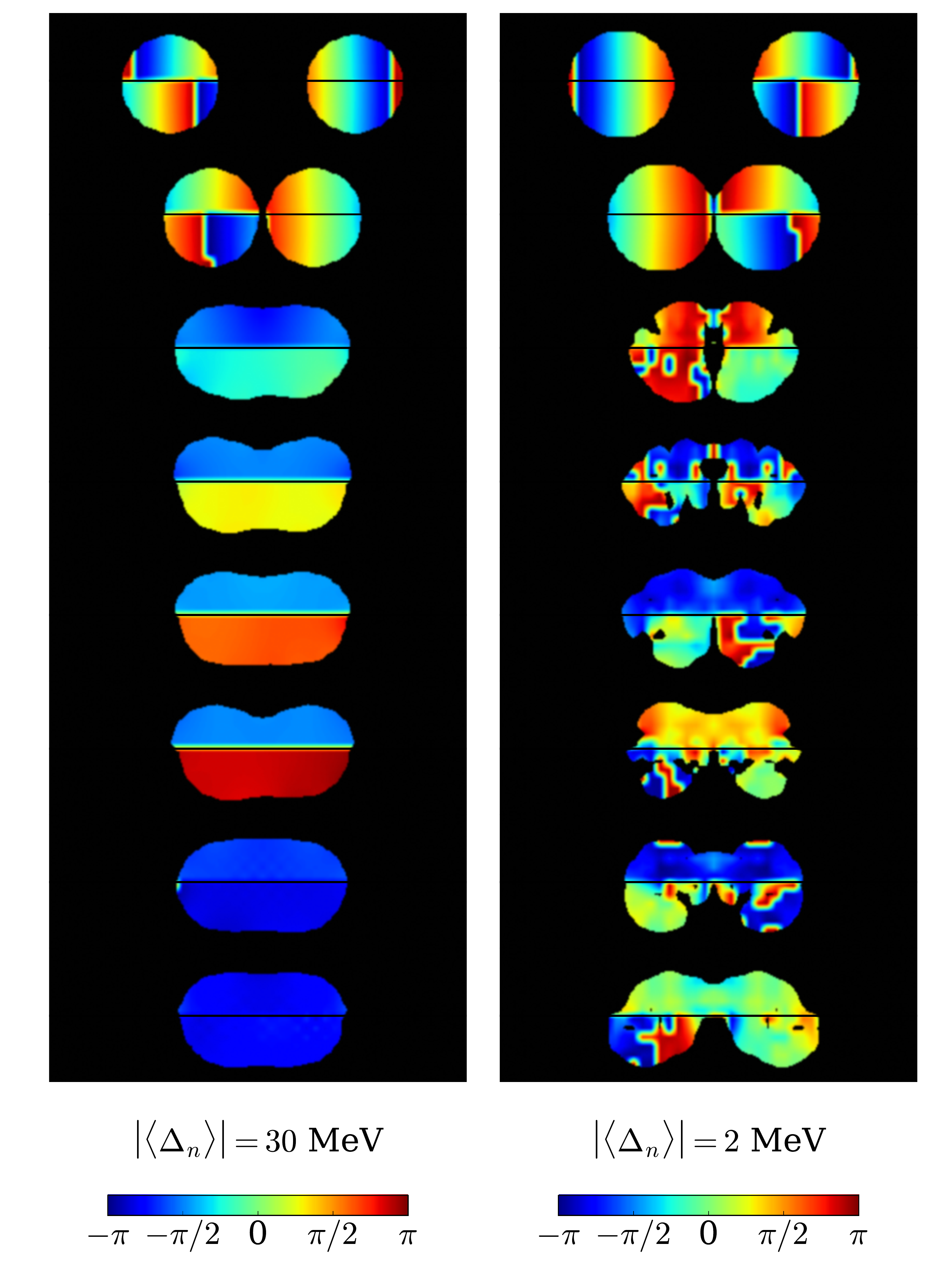}
\caption{\label{fig:coll} The evolution of the phase of the pairing
field (time runs top to bottom) in the head-on collision of
$^{120}$Sn+$^{120}$Sn~\cite{Bulgac:2017,Bulgac:2017c,Bulgac:2017d},
simulated with the phenomenological energy density functional SLy4 and
pairing as described in Ref.~\cite{Bulgac:2016}.  The right and the
left columns correspond to a realistic or artificially increased
pairing field strength respectively. The upper and lower half of each
frame corresponds to an initial phase difference between the two
initial pairing condensates of 0 and $\pi$ respectively. Even though
the pairing field magnitudes are constant before the colliding nuclei
come into contact, their phases change in time and space (first two
top frames), see Eq. \eqref{eq:1b}.  The phase locking of the pairing
field is clearly manifest after fusion in the left column, but absent
in the right column. }
\end{figure}

We observe the establishment of a common phase of the combined
condensate for large values of the coupling constant, which clearly
can be attributed to the phase rigidity in superfluids~\cite{
Ginzburg:1950, Ginzburg_rmp2004, Anderson_rmp1966, Anderson:1986,
Anderson:1984}.  While the two partners are in contact the phase of
the condensate becomes spatially constant over the entire system, and the
phase gets locked.  We illustrate the phase locking mechanism for both
Fermi and Bose superfluid systems: with the collision of two
superfluid nuclei described within the extension of TDDFT formalism to
superfluid fermionic systems ~\cite{Bulgac:2013b} by changing
the strength of the pairing correlations, see Fig.~\ref{fig:coll}, and
with the case of the collision of two BEC with relevant results in
SOM~\cite{Supplement}.

One can limit the analysis to a one-dimensional model as only matter,
momentum, and energy transfer between two colliding partners along the
line joining the two partners (which can rotate in space though) are
controlling most of the dynamics, similarly to the case of the
Josephson junction in the case of superconductors, when only dynamics
across the junction is typically analyzed.  In the absence of the
interaction ($g\equiv 0$), the GP equation is linear, and each wave
function $\psi_k({\bf r},t)$ satisfies independently the Schr\"odinger
equation
\beq i\hbar\dot{\psi}_k( {\bf r}, t) =
-\frac{\hbar^2}{2m}\triangle \psi_k({\bf r},t) + U({\bf r}, t)
\psi_k({\bf r},t), \nonumber 
\eeq 
and after the two potential wells have past each other each wave
function $\psi_k({\bf r},t)$ will split in between the two potential
wells.  Obviously, the linear combination of the wavefunctions
$\Psi({\bf r},t) =\psi_1({\bf r},t)+e^{i\alpha} \psi_2({\bf r},t)$,
which satisfies the same Schr\"odinger equation, depends on the
relative phase.  While for weak coupling $g$ the dynamics is
$\alpha$-dependent, when the strength of the interaction $g$ is
gradually increased, the dependence of the final outcome on the
relative phase $\alpha$ becomes weaker and weaker the stronger the
interaction gets, and the two cases $\alpha =0$ and $\alpha =\pi$ in
their final state become almost identical, see Figure~\ref{fig:coll}
for nuclei in 3D and SOM for bosons~\cite{Supplement}.
When the coupling constant is sufficiently large, the two boson clouds
penetrate each other and their final states are relatively little
affected irrespective of the value of $\alpha$ and both clouds emerge
with the initial number of particles practically unchanged and with
very small excitation energies as well~\cite{Supplement}.  The role of
the particle-particle interaction is to lead upon contact to a very
rapid phase locking between the two condensates after which the
properties of the final state depend very weakly on the phase
$\exp(i\alpha)$.  The strength of the interaction $g$ controls the
speed at which the information is transmitted throughout the cloud.
In the case of strong coupling, after the relatively short time needed
to send ``messages'' between the two partners, the properties of the
emerging final state are largely $\alpha$-independent and the two
clouds become completely entangled upon separation.  The total wave
function corresponds in this case to a coherent state in the particle
number difference $N_-=N_1-N_2$ and to a macroscopically entangled
state of two large objects.  This conclusion is in agreement with
Anderson's conjecture~\cite{Anderson:1986} concerning the inability of
an experimenter to recover the initial relative phase of the
condensates $\alpha$ after establishing the contact between the two
independently cooled liquid helium buckets from above
$T_\lambda$. This also it clarifies the content of Anderson's
conjecture, that only when the superfluid correlations are ``strong"
enough the role of the initial relative phase is erased. This is also
consistent with the generalized phase rigidity due to the term in GL
equation $n_s\hbar^2|{\bf \nabla}\phi|^2/2m$ (where $n_s$ is the
superfluid density) in the free energy of
superfluids~\cite{Ginzburg:1950, Ginzburg_rmp2004, Anderson_rmp1966,
Anderson:1986, Anderson:1984}, which is an emerging term, whose
presence and strength are dictated by the interactions, and which is
absent in non-interacting systems.

This dependence on $\alpha$ of the properties of the emerging
fragments in the case of ``weak" superfluid correlations reflects
particle number difference fluctuations between the two initial
partners, see also SOM~\cite{Supplement}.  The combined wave function
of two superfluid nuclei (with even particle numbers), depending on
two arbitrary gauge angles $\tau$ and $\alpha$, can be written
as~\cite{Bloch:1962} (here for simplicity for one kind of nucleons
only):
\bea \label{eq:2N}
| \Psi (\tau,\alpha)  &=& \prod_k [u_k + 
e^{i2\tau}e^{i2\alpha} v_k a^\dagger_{k} a^\dagger_{\overline{k}} ] \\
  &\times&            
  \prod_l  [u_l  + 
 e^{i2\tau} e^{- i2\alpha} v_l  a^\dagger_{l} a^\dagger_{\overline{l}} ] | 0 \rangle , 
\nonumber
\eea
where $k$ and $\overline{k}$ and $l$ and $\overline{l}$ denote pairs
of time-reversed states in the two nuclei and $u_{k,l}$ and $v_{k,l}$
being the corresponding amplitudes of the Bogoliubov-Valatin
quasiparticles.  Integrating $\Psi (\tau,\alpha)$ over $\tau$ with the
weight $e^{-i\tau N_+}$ will select the wave function with the total
particle number $N_+=N_1+N_2$. Integrating $\Psi (\tau,\alpha)$ over
$\alpha$ with the weight $e^{-iN_-\alpha}$ will select the exact
particle difference $N_-=N_1-N_2$ between the two nuclei.  In the case
of weak coupling an additional projection over the relative phase
$\alpha$ is required to ensure that the particle number difference
between the two initial partners has the expected value, namely
exactly zero ($\Delta N\equiv 0$) in the case of two identical nuclei,
see also SOM~\cite{Supplement}.  One can expect that total 
kinetic energy and fusion
rates distributions would become wider in case of superfluid colliding
nuclei.

When comparing our simulations of $^{240}$Pu
fission~\cite{Bulgac:2016} with realistic pairing interactions with
simulations in which the pairing field was artificially increased to
$\approx 3-4$ MeV~\cite{Bulgac:2017, Bulgac:2017c, Bulgac:2017d} we
observed a similar transition to a phase locking pattern: realistic
nuclear pairing strength is relatively weak, the phase locking does
not typically occur on the way from saddle-to-scission and the phase
and the magnitude of the pairing fields fluctuate strongly both in
space and time.  In the case of strong pairing~\cite{Bulgac:2017,
Bulgac:2017d, Bulgac:2017c}, even though the time from
saddle-to-scission is about ten times shorter, the evolution is almost
identical to the dynamics of an ideal or perfect fluid and the fission
fragments emerge strongly entangled. While one might naively expect a
faster rate of energy transfer from collective to intrinsic degrees of
freedom, the fluctuations of the pairing field are greatly suppressed
(due to larger gaps and larger critical velocities) and the evolving
fissioning nucleus stays cool.
 
In conclusion, we have established that the initial relative phase of
two colliding condensates plays an increasingly smaller role in the
case of strong interactions, when a phase locking over the entire
system is established fast (unless the entire system is very extended
and the signal propagation time is large as well), and after the
separation the final macroscopic (large) fragments emerge entangled.
  
We thank G.F. Bertsch for numerous discussions and P. Magierski,
K. Sekizawa, and G. Wlaz\l{}owski for commenting on a draft of the
manuscript. This work was supported in part by U.S. DOE Office of
Science Grant DE-FG02-97ER41014.  Calculations have been performed at
the OLCF Titan and at NERSC Edison.  This research used resources of
the Oak Ridge Leadership Computing Facility, which is a DOE Office of
Science User Facility supported under Contract DE-AC05-00OR22725 and
of the National Energy Research Scientific computing Center, which is
supported by the Office of Science of the U.S. Department of Energy
under Contract No. DE-AC02-05CH11231, and at Moonlight of the
Institutional Computing Program at Los Alamos National Laboratory.

 \bibliography{local}


\section{Supplemental Online Material}
  
%
%
\vspace{0.75cm}

\begin{figure}
\includegraphics[clip,width=0.45\columnwidth]{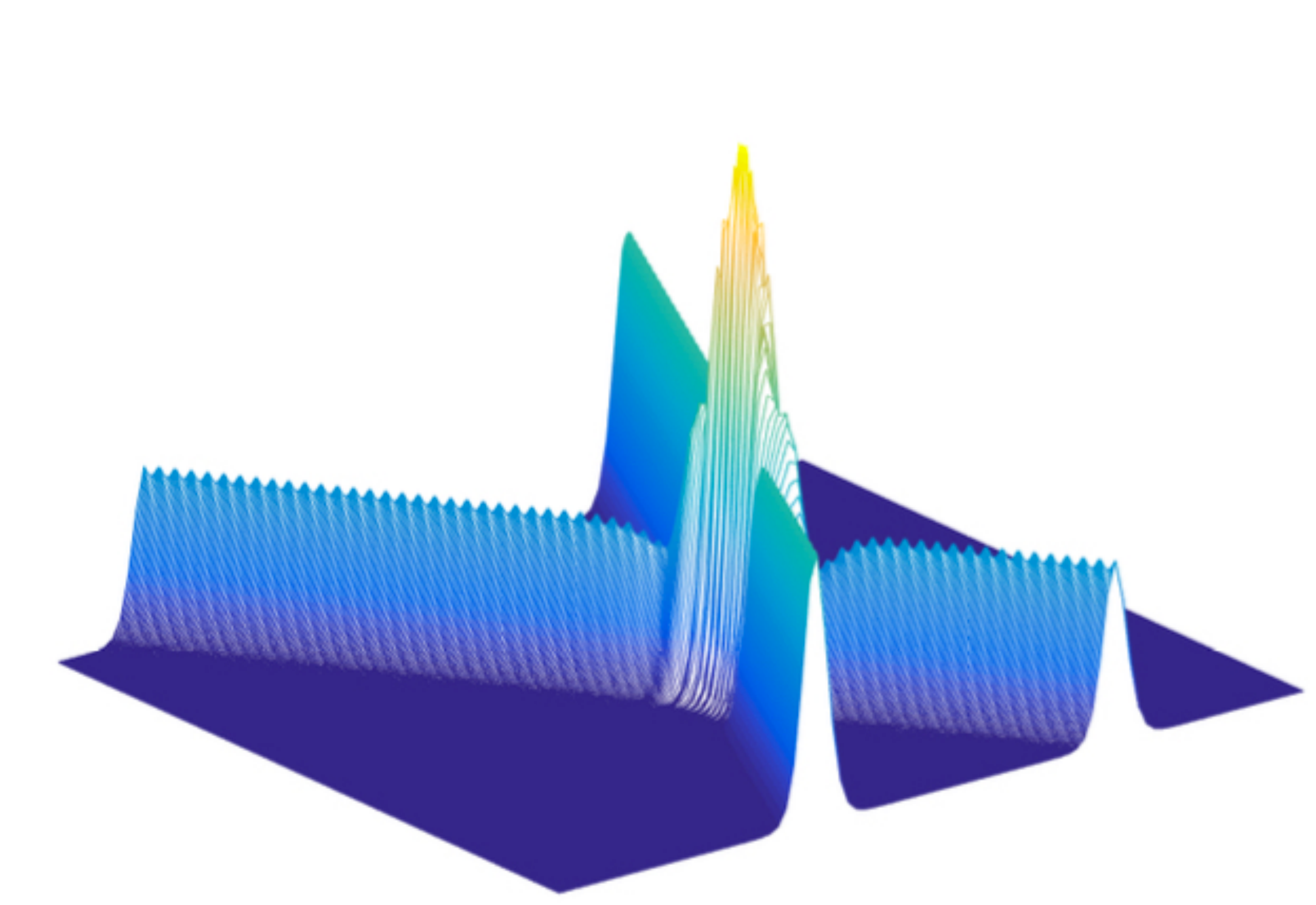}
\includegraphics[clip,width=0.45\columnwidth]{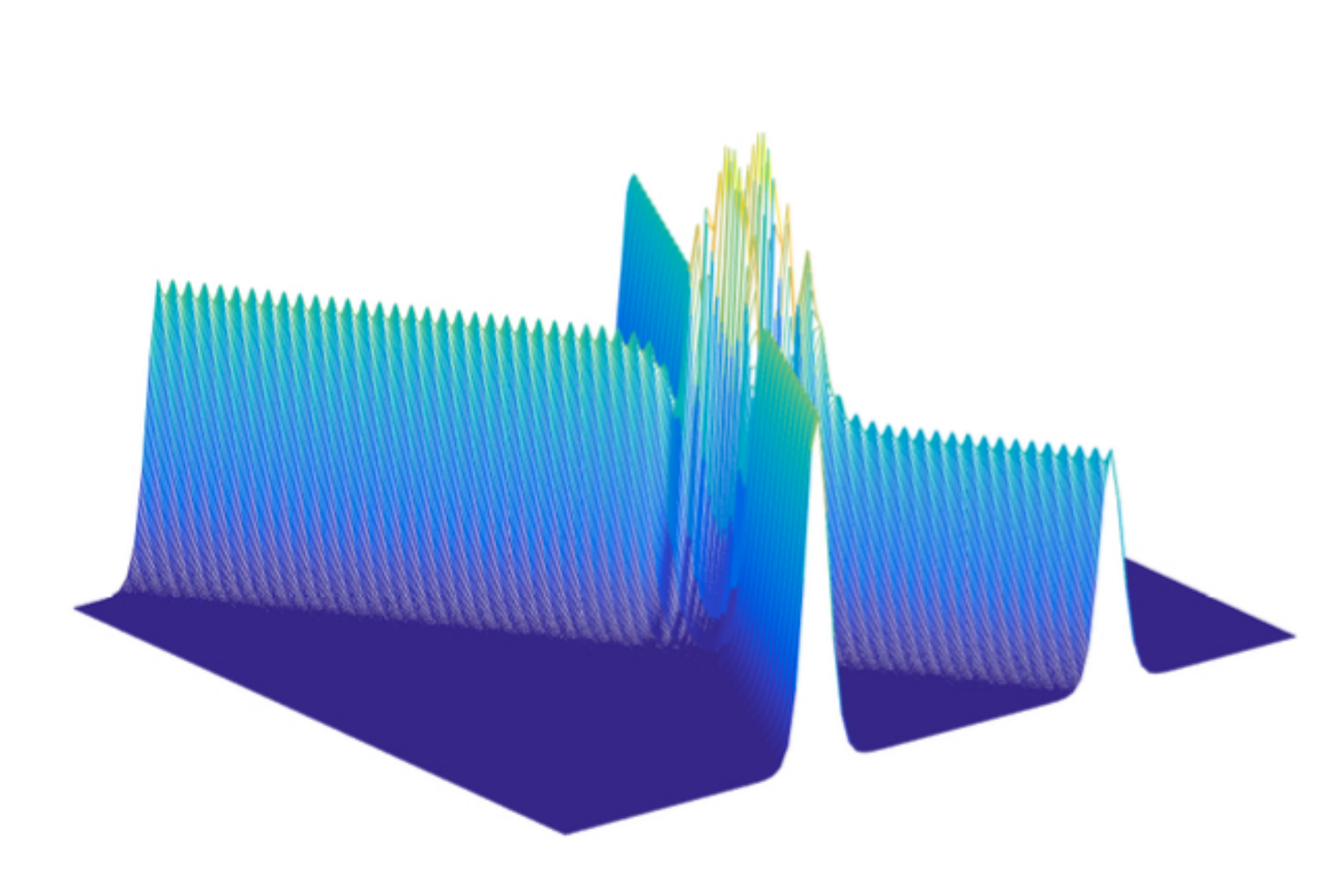}
\includegraphics[clip,width=0.45\columnwidth]{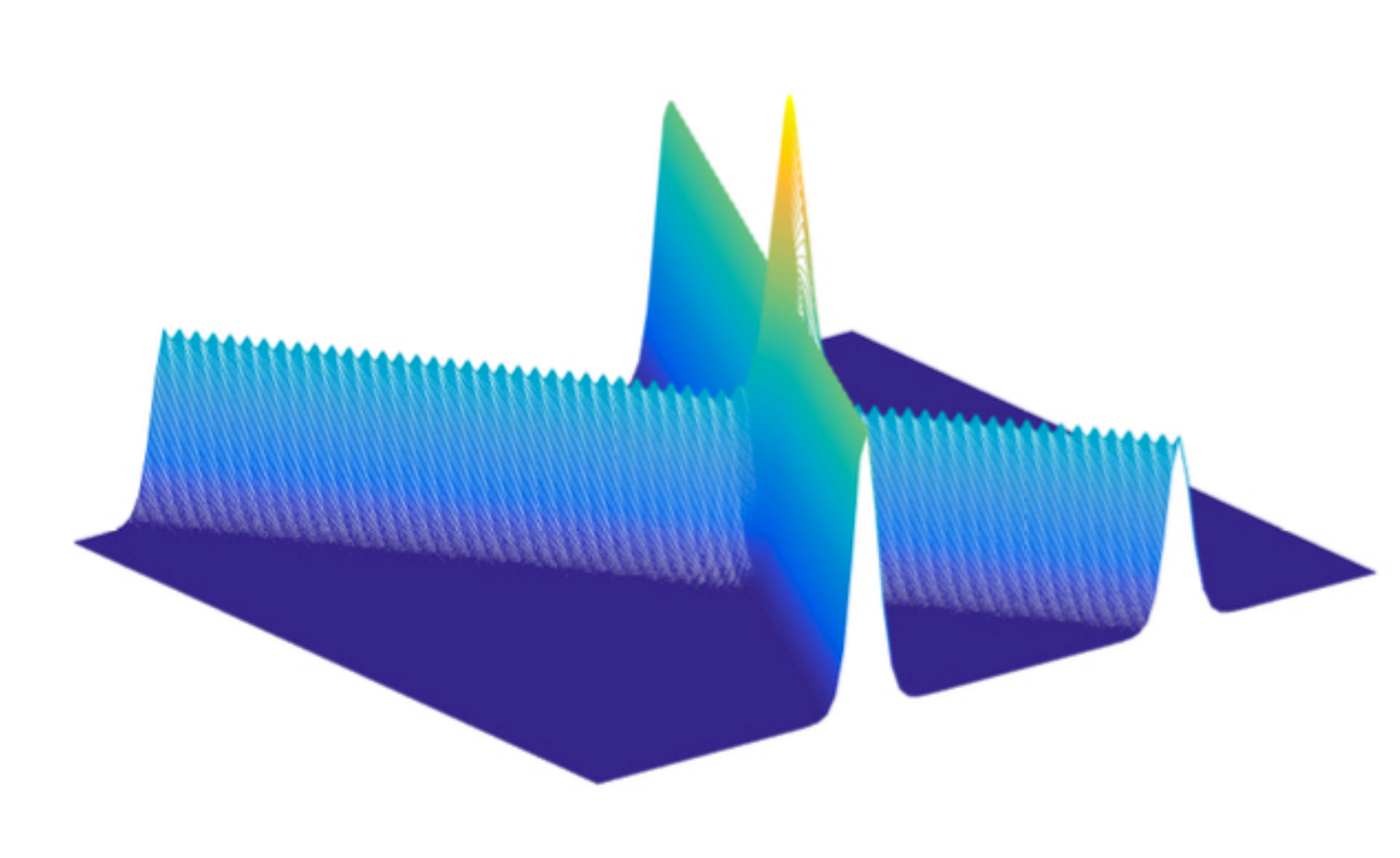}
\includegraphics[clip,width=0.45\columnwidth]{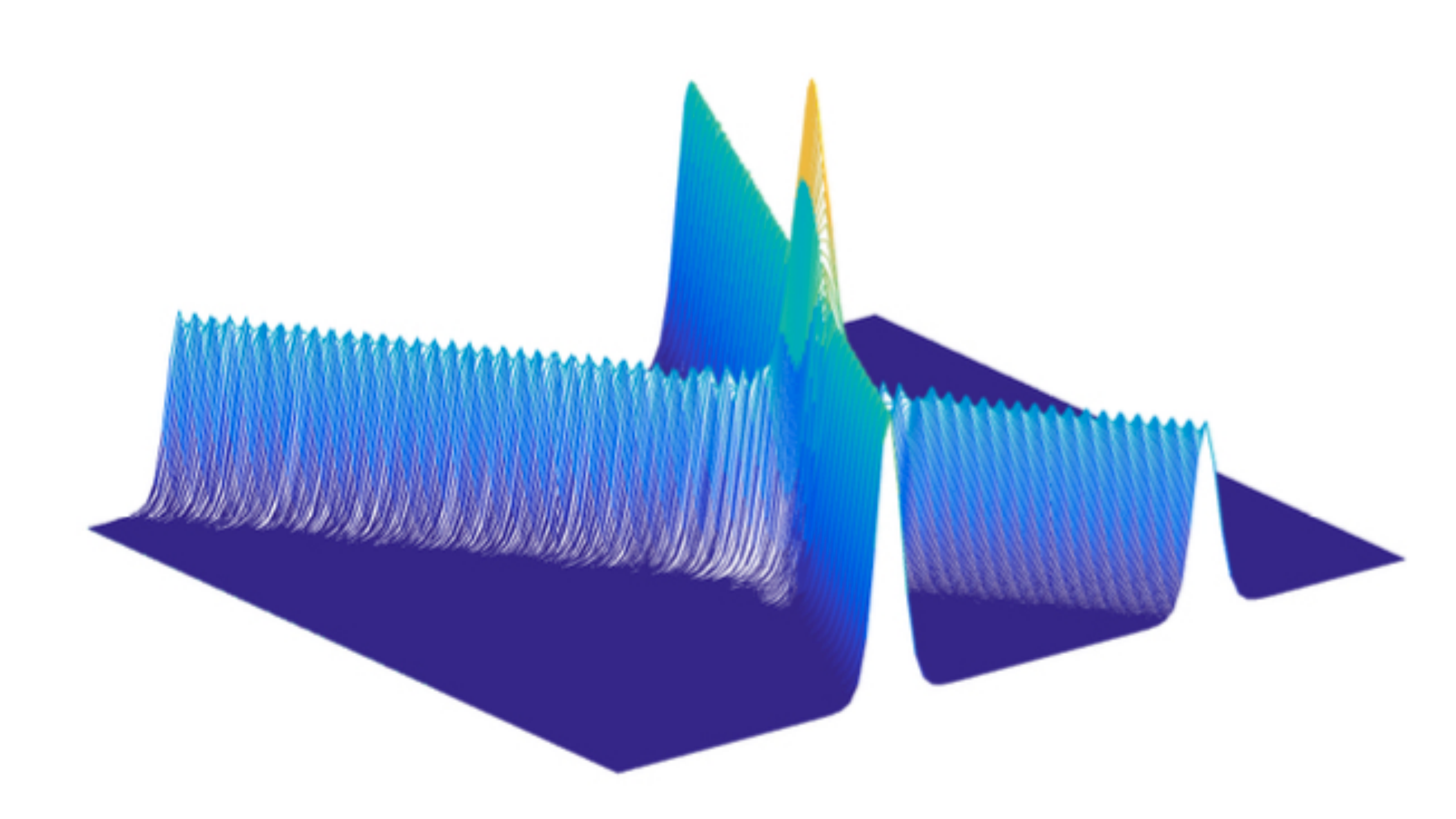}
\includegraphics[clip,width=0.45\columnwidth]{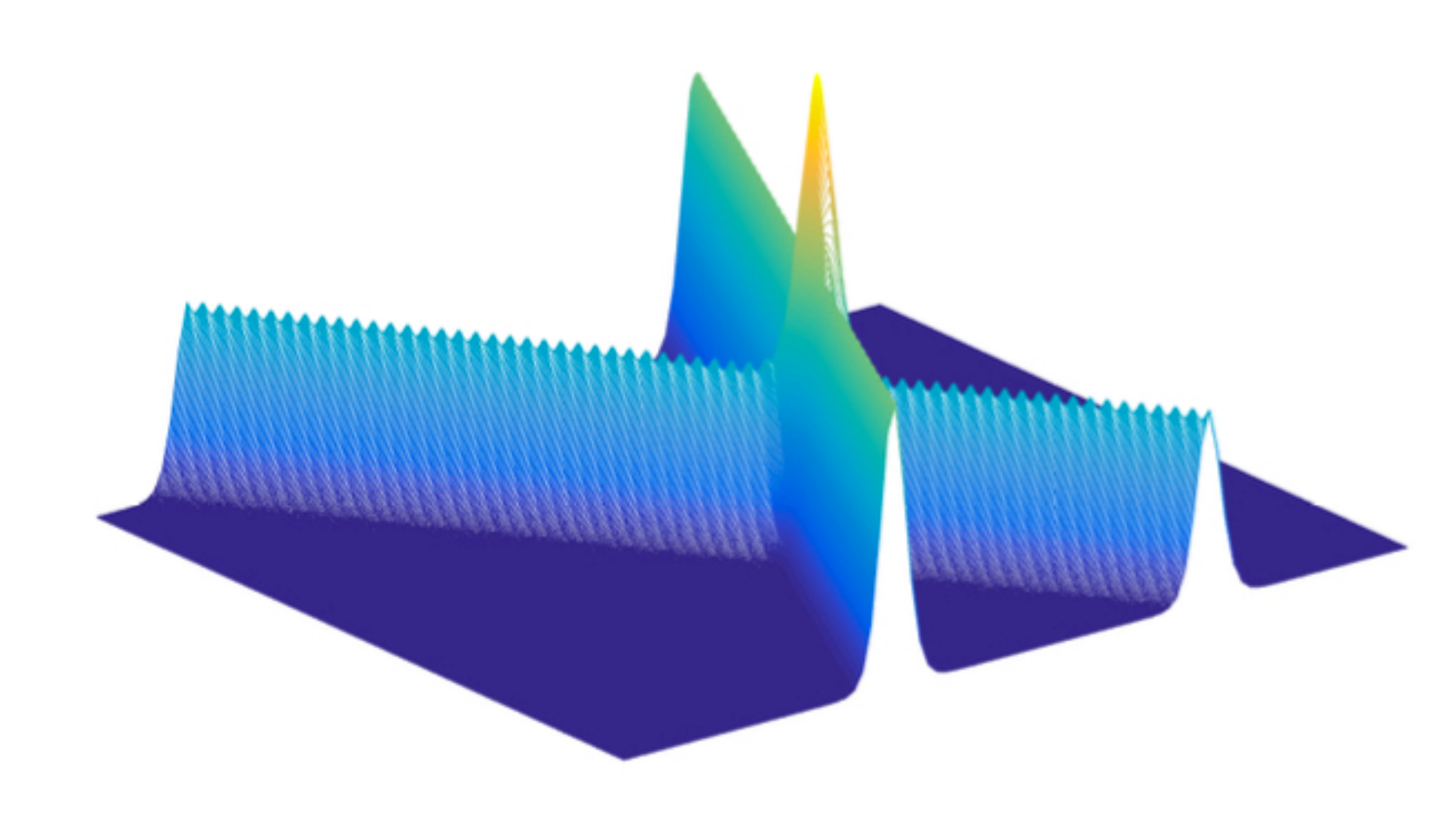}
\includegraphics[clip,width=0.45\columnwidth]{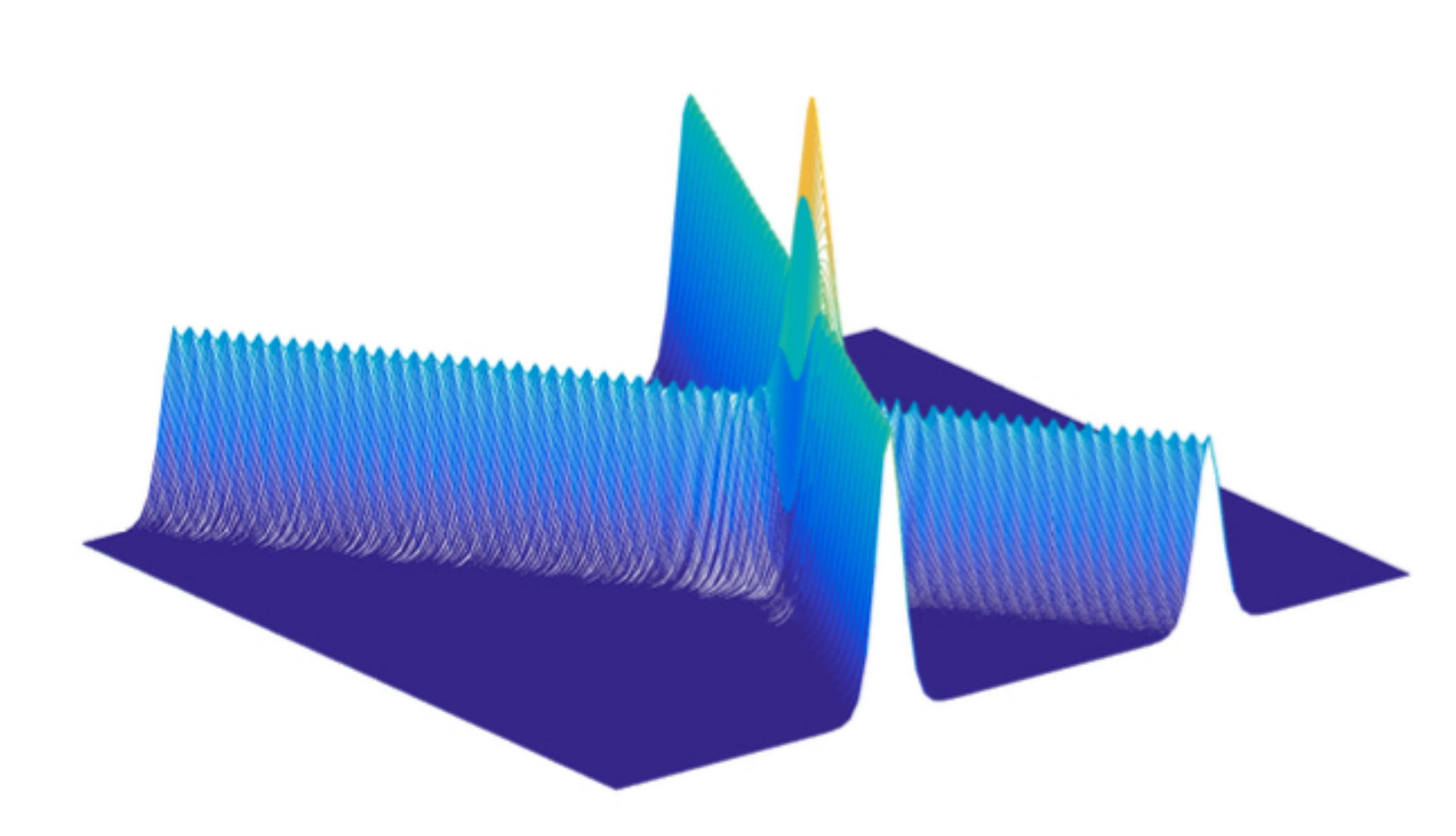}
\caption{\label{fig:bec_coll} The space-dependence evolution of
$|\Psi(x,t)|^2$ for two colliding BEC clouds trapped in slightly
modified Posh-Teller potentials $V_k({\bf r}-{\bf r}_k- {\bf u}_k t)$, with $N_1=120$,
$N_2=80$, and for interaction strengths $g=0, 0.5,  1$ (upper, middle, 
and lower rows respectively) and for two values of the mixing
parameter $\alpha=0,\pi$ (left and right columns respectively).  The
time runs from lower corner to the left and the coordinate $x$ from
the same corner to the right. In the case of weak coupling the dynamics 
is qualitatively similar to the non-interacting case (upper row). In the case $g\neq 0$ 
the two clouds behave as they are almost fully transparent.}
\end{figure}

We created two initial BEC clouds in a modified Posh-Teller external potential (units $\hbar=2m=1$)
\bea
&&V_k(x)= -\frac{s_k(s_k+1)}{\cosh^2(x)} - g\frac{N_kn_k}{\cosh^{2s_k}(x)},\\
&&\int dx \frac{1}{\cosh^{2s_k}(x)}=n_k, \quad k=1,2
\eea
and where $N_K$ is the number of bosons in each cloud. With this choice the state
\beq
\phi_k(x)=\frac{\sqrt{n_k}}{\cosh^{s_k}(x)}
\eeq
is a stationary solution of the GP equation
\bea
&&-\frac{d^2\phi_{k}(x)}{dx^2}+V_k(x)\phi_k(x)+gN_k|\phi^k(x)|^2\phi_k(x)\\\nonumber
&&=-s_k(s_k+1)\phi_{k}(x)
\eea
with $s_k\geq 1$. As described in the main text, we initialize two spatially separated BEC clouds 
with particle numbers $N_1$ and $N_2$ and collide them with various relative velocities, 
chosen below the critical velocity, see Figure~\ref{fig:bec_coll}. 

\begin{figure}
\includegraphics[clip,width=0.99\columnwidth]{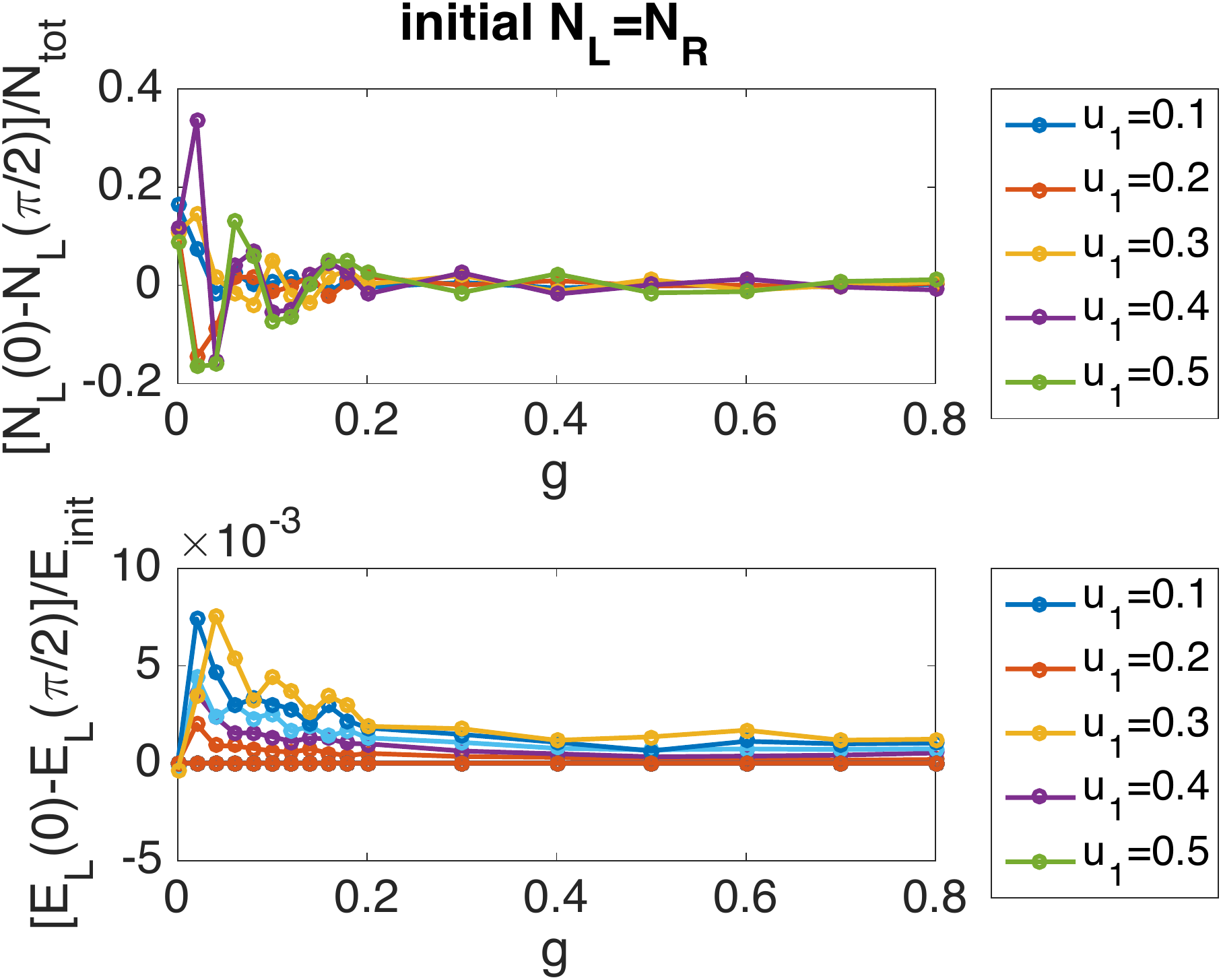}
 \caption{\label{fig:nl}  Two identical clouds with $N_1=N_2=100$ particles
 were collided with various relative velocities ($u_1$ is the initial velocity 
 of the first cloud) and for varying coupling constants $g$ and a relative 
 initial phase of the two BEC condensates $\pi/2$. The difference between 
 the particle numbers after the collision  and the difference in the  excitation
  energy for a relative phase of the two condensates $0$ and $\pi/2$ in the l
  eft cloud after the collision are shown in the upper and lower panels respectively.  }
\end{figure}

\begin{figure}
\includegraphics[clip,width=0.99\columnwidth]{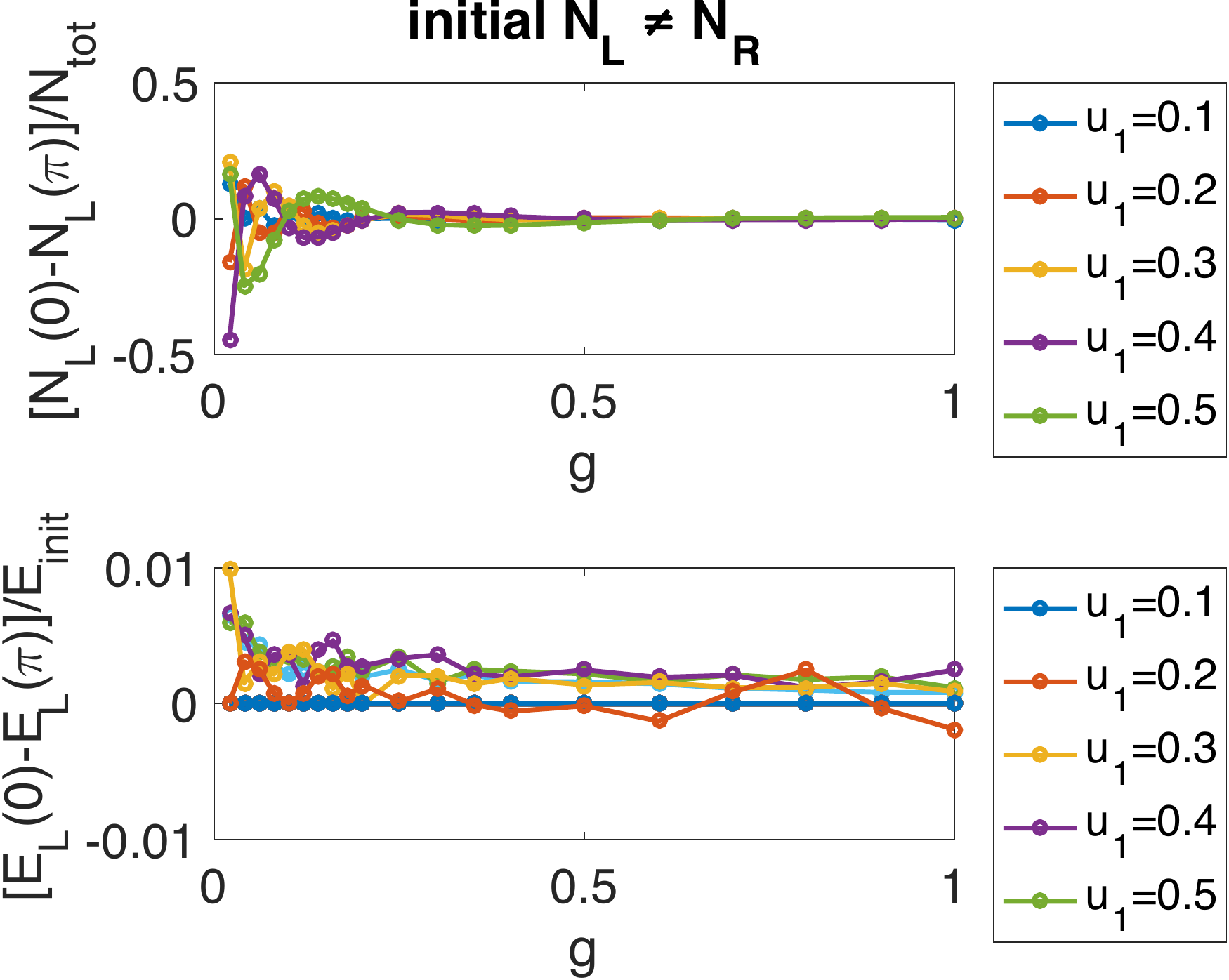}
 \caption{\label{fig:nl1}  Two non-identical clouds with $N_1=125$ 
 and $N_2=75$ particles
 were collided with various relative velocities ($u_1$ is the initial 
 velocity of the first cloud) and for varying coupling constants $g$ 
 and a relative initial phase of the two BEC condensates $\pi$. The 
 difference between the particle numbers after the collision  for 
 relative phase $0$ and $\pi$,  and the difference in the  excitation 
 energy for relative phase $0$ and $\pi$ in the left cloud after 
 the collision are shown in the upper and lower panels respectively.  }
\end{figure}

In Figures~\ref{fig:nl} and \ref{fig:nl1} 
we show the differences between the particle numbers and excitation energies in a cloud for different 
initial relative phases of the two colliding BECs. With increasing coupling 
strength the clouds show less and less dependence of the final state properties 
on the initial relative phase difference of the condensates. The transition from one 
regime, where the dependence of the properties of the final fragments depend strongly 
on the initial relative phase between the two condensates, to the regime, where this 
dependence is almost absent, is gradual. The largest differences are observed for ``small'' coupling constants.

According to the general Hohenberg-Kohn
theorem there is a
one-to-one correspondence between the many-body wave function and the
(generalized) density and thus in principle the many-body wavefunction
can be determined from the (generalized) density $\Psi[\rho(\alpha)]$
(the dependence on the
phase $\alpha$ is explicit). Since for superfluid systems the density
$\rho(\alpha)$ describes a system with an average particle number
only, the corresponding many-body wavefunction $\Psi[\rho(\alpha)]$
describes a system with only an average particle number.  Any
observable can be in principle evaluated using this many-body wave
function, or in other words, for any many-body observable there exists
a corresponding density functional as well, $O[\rho] = \langle
\Psi[\rho] |\hat{O}|\Psi[\rho]\rangle $.  Most of observable of
interest $\hat{O}$ conserve the particle numbers, i.e.  $\hat{O}$ commutes
with the particle projector $\hat{P}_N=\hat{P}_N^2$.  $\hat{P}_N$
can project either on the entire system or on each incident or emerging nucleus
separately.  The value of an observable for a fixed particle number
$N$ can be evaluated as follows 
\bea
O_N(\alpha)&=& \langle \Psi_N(\alpha) |\hat{O}|\Psi_N(\alpha)\rangle , \label{eq:N} \\
|\Psi_N(\alpha)\rangle &=&    \frac{\hat{P}_N|\Psi[\rho(\alpha)] 
\rangle}{\sqrt{\langle \Psi[\rho(\alpha)]  | \hat{P}_N|\Psi[\rho(\alpha)] \rangle}}.  \label{eq:PN}
\eea
where $|\Psi_N(\alpha)$ is the projected many-body 
wavefunction, which may retain a dependence on
the phase $\alpha$, see discussion in the main text in connection to  and Ref. [13],
P.W. Anderson (1986).  The crucial question is: Is the projected
many-body function $|\Psi_N(\alpha)\rangle $ really
$\alpha$-dependent?

In the
cases discussed in the main text  with weak coupling  an additional
projection over the relative phase $\alpha$ is required to ensure that
the particle number difference between the two identical initial partners has
the expected value, namely exactly zero ($\Delta N\equiv 0$) in the
case of two identical nuclei. 

In the case of two bosons the wave function is
\bea \label{eq:2B}
\Psi( {\bf r}_1, {\bf r}_2 ,t)&=&[\psi_1( {\bf r}_1 ,t)+e^{i\alpha}\psi_2( {\bf r}_1 ,t)] \\
&\times&[\psi_1( {\bf r}_2, t)+e^{i\alpha}\psi_2( {\bf r}_2, t)], \nonumber
\eea
\bea
&=& \psi_1( {\bf r}_1 ,t)\psi_1( {\bf r}_2, t) \nonumber \\
&+& e^{i\alpha} [\psi_1( {\bf r}_1 ,t)\psi_2( {\bf r}_2 ,t)+\psi_1( {\bf r}_2, t)\psi_2( {\bf r}_1, t)]  \nonumber \\
&+&  e^{2i\alpha} \psi_2( {\bf r}_1 ,t)\psi_2( {\bf r}_2, t)  \nonumber
\eea
and only the part
\beq
e^{i\alpha} [\psi_1( {\bf r}_1 ,t)\psi_2( {\bf r}_2 ,t)+
\psi_1( {\bf r}_2, t)\psi_2( {\bf r}_1, t)]
\eeq
will describe correctly 
the initially two separated particles. The same applies to a similarly extracted 
component in the case of many bosons. 

In the case when the phase locking of the two condensates is
established fast, this kind of projection is largely unnecessary. In
nuclear systems with realistic pairing coupling strength  the outcome of the collisions depends on the
relative phase $\alpha$ 
and observables should be projected on the appropriate particle 
number difference $\Delta N$, in order to remove spurious fluctuations
in the particle number differences between the partners. In practice, in 
the case of realistic nuclear pairing interactions, one might opt to evaluate
the average of  $\langle \Psi(\alpha) |\hat{O}|\Psi(\alpha)\rangle$
over the phase $\alpha$,  instead of preferably evaluating the projected
on $\Delta N$ value
$\langle \Psi_{\Delta N}(\alpha) |\hat{O}|\Psi_{\Delta N}\alpha)\rangle$, 
see Eqs.(\ref{eq:N}, \ref{eq:PN}). These 
two quantities might be turn out to be similar in practice. 
 
In the case of the collision of two nuclei, illustrated in Fig. 1 in the main text,  
we have followed the same calculational procedure 
as in Ref. [32] cited in the main text, using for the 
normal part of the nuclear energy functional SLy4 with no omissions, and we 
have only varied the strength of the pairing coupling constant from a realistic 
value for nuclei to a value which brings the paring correlations for neutrons in 
the two $^{120}$Sn nuclei to a regime close to the unitary Fermi gas, when 
the pairing gap is of the order of the Fermi energy.  We have initiated the 
two nuclei in their ground state, separated by a relative large distance, 
after which we have accelerated them towards each other.

\end{document}